\documentclass[%
	aip,%
	jcp,%
	amsmath,amssymb,%
	preprint,%
	a4paper,%
]{revtex4-1}

\usepackage{graphicx}
\usepackage{lipsum}
\usepackage{color}
\usepackage{upgreek}
\usepackage[normalem]{ulem}

\renewcommand{\figurename}{Fig.}
\renewcommand{\tablename}{Table}

\begin{document}


\title{Vacuum-ultraviolet frequency-modulation spectroscopy} 



\author{U.~Hollenstein}
\affiliation{Laboratorium f\"ur Physikalische Chemie, ETH Z\"urich, 8093 Z\"urich, Switzerland}
\author{H.~Schmutz}
\affiliation{Laboratorium f\"ur Physikalische Chemie, ETH Z\"urich, 8093 Z\"urich, Switzerland}
\author{J.~A.~Agner}
\affiliation{Laboratorium f\"ur Physikalische Chemie, ETH Z\"urich, 8093 Z\"urich, Switzerland}
\author{M.~Sommavilla}
\affiliation{Laboratorium f\"ur Physikalische Chemie, ETH Z\"urich, 8093 Z\"urich, Switzerland}
\author{F.~Merkt}
\affiliation{Laboratorium f\"ur Physikalische Chemie, ETH Z\"urich, 8093 Z\"urich, Switzerland}


\date{\today}

\begin{abstract}
Frequency-modulation (FM) spectroscopy has been extended to the vacuum-ultraviolet (VUV) range of the electromagnetic spectrum. Coherent VUV laser radiation is produced by resonance-enhanced sum-frequency mixing ($\nu_{\mathrm{VUV}}=2\nu_{\mathrm{UV}}+\nu_2$) in Kr and Xe using two near-Fourier-transform-limited laser pulses of frequencies $\nu_{\mathrm{UV}}$ and $\nu_2$. Sidebands generated in the output of the second laser ($\nu_2$) using an electro-optical modulator operating at the frequency $\nu_{\mathrm{mod}}$ are directly transfered to the VUV and used to record FM spectra. Demodulation is demonstrated both at $\nu_{\mathrm{mod}}$ and $2\nu_{\mathrm{mod}}$. The main advantages of the method 
compared to VUV absorption spectroscopy is its background-free nature, the fact 
that its implementation using table-top laser equipment is straightforward and that it can be used to record VUV absorption spectra of cold samples in skimmed supersonic beams simultaneously with laser-induced-fluorescence and photoionization spectra. To illustrate these advantages we present VUV FM spectra of Ar, Kr, and N$_2$ in selected regions between 105\,000\,cm$^{-1}$ and 122\,000\,cm$^{-1}$. 
\end{abstract}


\maketitle 

\section{Introduction}\label{sec:intro}
\begin{figure}[!ht]\centering
	\includegraphics[scale=1.0]{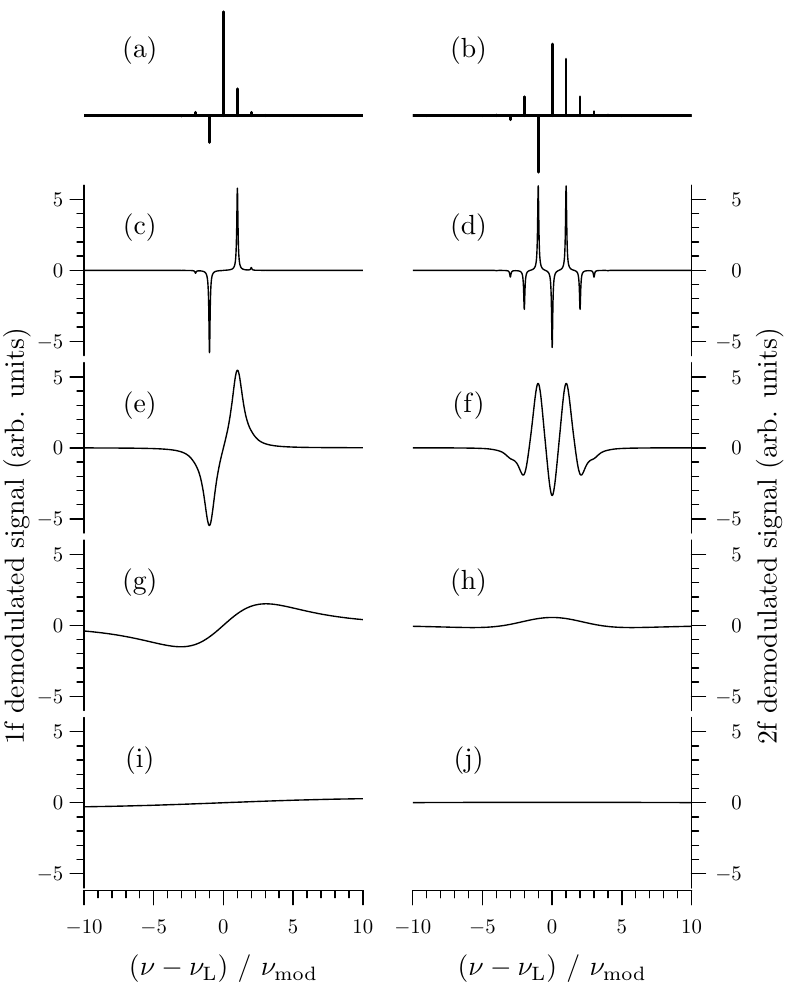}%
	\caption{(a): Amplitude spectrum of a laser of frequency $\nu_{\mathrm L}$ modulated at frequency $\nu_{\mathrm{mod}}$ with a modulation index $\beta=0.5$. The sign indicates the phase of the corresponding sideband. (c), (e), (g) and (i): Simulations of demodulated signals for Lorentzian line profiles with full width at half maximum of $\varGamma=\frac1{10}\nu_{\mathrm{mod}}$, $\nu_{\mathrm{mod}}$, $10\nu_{\mathrm{mod}}$ and $50\nu_{\mathrm{mod}}$, respectively, and phase $\phi=0$. 
	(b): Amplitude spectrum of a laser of frequency $\nu_{\mathrm L}$ modulated at frequency $\nu_{\mathrm{mod}}$ with a modulation index $\beta=1.25$. (d), (f), (h) and (j): Simulations of the demodulation signals at a phase of $\phi=0$ obtained at the same values of $\varGamma$.  
	}\label{fig:mod-spec}
\end{figure}

Frequency-modulation (FM) spectroscopy is a very sensitive and powerful method to record atomic and molecular laser spectra in the gas phase \cite{bjorklund80a,bjorklund81a,lenth81a,gallagher82a,tran84a,janik85a,eyler96a}. Originally, the method was introduced to exploit the advantages of narrow-band lasers for spectroscopic investigations in the infrared (IR) and the visible (VIS) ranges of the electromagnetic spectrum. Eyler and coworkers later applied the method in combination with nonlinear frequency-upconversion methods to extend the wavelength range to the UV down to 214.5\,nm and achieved an absorption sensitivity of $8\cdot10^{-4}$ in this range.\cite{eyler96a} The high sensitivity results from the background-free nature of the detection of the absorption signals. FM spectroscopy relies on the modulation at frequency $\nu_{\mathrm{mod}}$ of a narrow-band laser operated at frequency $\nu_{\mathrm L}$, resulting in an amplitude spectrum schematically depicted in \figurename~\ref{fig:mod-spec} for the modulation indices $\beta=0.5$ and 1.25 relevant for the present investigation ($\beta$ defines the sideband-to-carrier ratio, see Ref.~\onlinecite{supplee94a} for additional details). Maximal sensitivity is reached when the modulation frequency is larger than the intrinsic linewidth of the transitions. This property prompted Janik {\it et al.} \cite{janik85a} to demodulate the signals at a frequency $2\nu_{\mathrm{mod}}$ instead of $\nu_{\mathrm{mod}}$ to improve the contrast in the case of a linewidth $\varGamma$ comparable to $\nu_{\mathrm{mod}}$.

\begin{figure*}\centering
	\includegraphics[width=\linewidth]{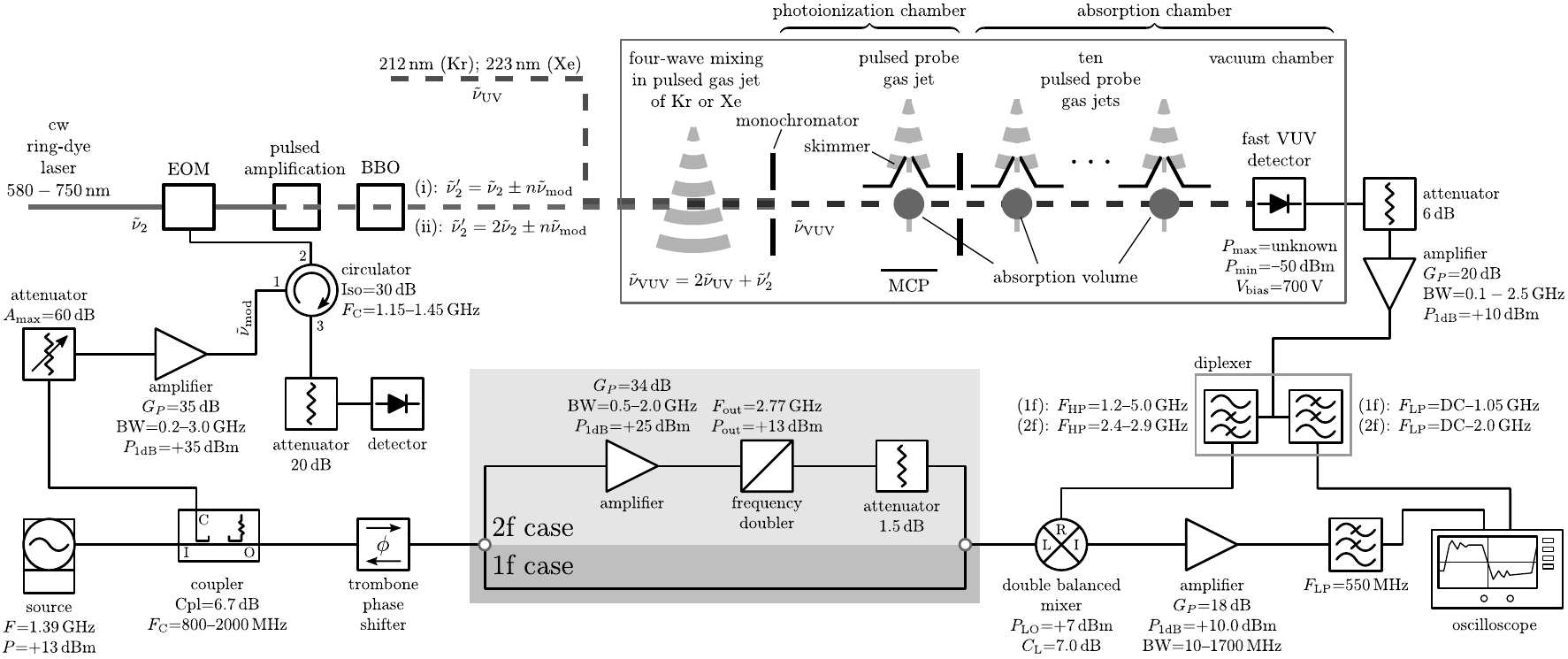}%
	\caption{Schematic representation of the experimental setup. Upper part: optical setup of the VUV absorption experiment. Lower part: modulation/demodulation components (see text for details). \label{fig:expscheme}}%
\end{figure*}%

In this article, we present an extension of FM spectroscopy to the vacuum-ultraviolet (VUV, $\lambda<200\,\mathrm{nm}$) range of the electromagnetic spectrum and explore the sensitivity of demodulation at both $\nu_{\mathrm{mod}}$ and $2\nu_{\mathrm{mod}}$. High-resolution absorption spectroscopy below 200\,nm is notoriously difficult. In early high-resolution spectroscopic work in the VUV, the absorption spectra were recorded using high-pressure lamps in combination with large VUV monochromators \cite{tanaka69a,chupka69a,yoshino70a,herzberg72a,berkowitz79a}, enabling resolution of about 0.5\,cm$^{-1}$ between 10\,eV and 20\,eV. By replacing the high-pressure lamps by synchrotron radiation, spectroscopic measurements could be extended beyond 20\,eV. VUV radiation from synchrotron sources, however, also needs to be monochromatized, which limits the bandwidth of the radiation to at best $0.1\,\mathrm{cm}^{-1}$.\cite{nahon01a} VUV Fourier-transform absorption spectroscopy using synchrotron radiation was recently extended to the range below 105\,nm and offers the multiplex advantage, but so far the best resolution achieved with this method is $0.33\,\mathrm{cm}^{-1}$.\cite{deoliveira09a} Pulsed VUV laser systems based on four-wave mixing enable a higher resolution (better than 200\,MHz, see Refs. \onlinecite{cromwell89a,hinnen98a,hollenstein00a,trickl07a}), but the large pulse-to-pulse fluctuations resulting from the nonlinearity of the VUV generation process limits the sensitivity of absorption measurements. Consequently, only few VUV-laser-absorption spectra of atoms and molecules in supersonic beams have been reported.\cite{hinnen98a,softley87a,trickl89a,trickl07a} 
Supersonic beams enable one to cool the internal degrees of freedom of molecules and to reduce spectral congestion and Doppler broadening, however, at the cost of a reduced column density.
To improve the low sensitivity resulting from the large pulse-to-pulse fluctuations of the VUV laser radiation, Sommavilla \emph{et al.} \cite{sommavilla02a,sommavilla16a} have used a dispersion grating and exploited the beam diffracted in the negative first order to normalize the VUV laser intensity pulse by pulse and were able to reliably measure relative changes of the transmission of $10^{-4}$, which is sufficient to record VUV absorption spectra of molecules in cold supersonic beams.

Although (1+1$'$) resonance-enhanced multiphoton-ionization (REMPI) spectroscopy can be used to record VUV spectra with high sensitivity, the line intensities may be reduced, compared to an absorption spectrum, by predissociation. Here, we present an alternative method to record VUV absorption spectra with high sensitivity that relies on FM techniques. Tunable VUV radiation is produced from the output of two Fourier-transform-limited pulsed lasers (pulse length $5\,\mathrm{ns}$, obtained by pulse amplification of cw single-mode ring laser radiation) by two-color resonance-enhanced four-wave mixing ($\tilde{\nu}_{\mathrm{VUV}}=2\tilde{\nu}_{\mathrm{UV}}+\tilde{\nu}_2$) in Kr or Xe. The modulation of the VUV laser frequency is achieved by generating sidebands on the output of the second laser ($\tilde{\nu}_2$) using an electro-optical modulator. These sidebands are automatically transferred to the VUV, because the four-wave-mixing process is linearly dependent on the intensity of the second laser.

\section{Experimental Setup and Procedure}\label{sec:expsetup}
The experimental setup used in this work is depicted schematically in \figurename~\ref{fig:expscheme} and consists of a laser system (Section \ref{sec:laser}) and vacuum chambers where the absorption experiments are carried out on cold gaseous samples in skimmed supersonic beams. Critical components for the success of the experiment such as the (de)modulation setup and the home-built VUV photodetector are described in Sections \ref{sec:mod-demod} and \ref{sec:detector}, respectively. Section \ref{sec:signaldetection} presents the different detection schemes used to characterize the VUV absorption and Section \ref{sec:analysis} provides details of the analysis of the lineshapes.
\subsection{VUV Laser System}\label{sec:laser}
The near-Fourier-transform-limited VUV laser source \cite{hollenstein00a,sommavilla02a} used in the present work is depicted schematically in the upper part of \figurename~\ref{fig:expscheme}. VUV radiation at the wave number $2\tilde\nu_{\mathrm{UV}}+\tilde\nu_2$ is produced in a resonance-enhanced four-wave mixing process in atomic krypton and xenon using the two-photon resonances Kr (4p)$^5$ 5p[1/2]$_0$ $\leftarrow$ Kr (4p)$^6$ $^1$S$_0$ at $2\tilde\nu_{\mathrm{UV}}=94\,092.863$\,cm$^{-1}$ (Ref. \onlinecite{saloman07a}) and Xe (5p)$^5$ 6p$'$[1/2]$_0$ $\leftarrow$ Xe (5p)$^6$ $^1$S$_0$ at $2\tilde\nu_{\mathrm{UV}}= 89\,860.015$\,cm$^{-1}$ (Ref. \onlinecite{NIST2016a}). Fourier-transform-limited pulses at the wave numbers $\tilde\nu_1=\frac13\tilde\nu_{\mathrm{UV}}$ and $\tilde\nu_2$ are generated by pulsed amplification of the continuous-wave single-mode output of two ring dye lasers (Coherent 699-21 and 899-29, output power of about 500\,mW) using dye cells pumped by the second harmonic of a pulsed, injection-seeded Nd:YAG laser (Spectra-Physics, Quanta Ray Lab 170, pulse length of about 8\,ns, repetition rate 25\,Hz). To generate the desired wave number $\tilde\nu_{\mathrm{UV}}$, the amplified laser radiation is up-converted to $\tilde\nu_{\mathrm{UV}}=3\tilde\nu_1$ using two successive $\upbeta$-barium-borate crystals (BBO), see Ref. \onlinecite{hollenstein00a} for details. The beam with wave number $\tilde\nu_2$ is the pulsed-amplified fundamental output of the second ring dye laser or its second harmonic. As long as the modulation index of the fundamental output remains below 0.5, the doubled output is characterized by approximately the same modulation index as the fundamental because the sidebands are too weak to be efficiently frequency doubled. The two two-photon resonances listed above enable the generation of VUV radiation over a broad spectral range of about 60\,000\,cm$^{-1}$ to 135\,000\,cm$^{-1}$.

The sum-frequency VUV laser radiation with wave number $\tilde\nu_{\mathrm{VUV}}=2\tilde\nu_{\mathrm{UV}}+\tilde\nu_2$ is separated from the fundamental laser beams and beams generated through other nonlinear processes in a vacuum monochromator. The separation is achieved with a toroidal grating, which also refocusses the VUV radiation at the exit slit of the monochromator chamber.\cite{merkt98a} The detected VUV-laser beam intersects one or more pulsed skimmed supersonic beams of the sample gas at right angles in the absorption chambers. The VUV-laser pulses have a duration (full width at half maximum) of approximately $2.5$\,ns and a bandwidth of approximately 300\,MHz.

The absorption chambers consist of two separate differentially pumped regions. The probe-gas beams are generated by supersonic expansion using pulsed valves (Parker, General Valve, Series 9, valve orifice diameter 1\,mm) operated at a stagnation pressure of about 2\,bar. The opening time of the valves is typically 200\,$\upmu$s. In the first region, referred to as photoionization chamber below, a single probe-gas beam is collimated at a distance of 2\,cm from the valve orifice by a skimmer (Beam Dynamics, orifice diameter 2\,mm). Photoexcitation takes place on the axis of a linear time-of-flight photoionization mass spectrometer with which photoionization spectra can be recorded after the photoions are extracted with a pulsed electric field towards a microchannel-plate (MCP) detector. With this detector, the fluorescence induced by the VUV radiation can also be monitored. In the second region, referred to as absorption chamber below, up to ten nozzles located 2.0\,cm above their respective skimmers (orifice diameter 1\,mm) are used.\cite{sommavilla16a} The transmitted VUV intensity is detected using a home-built VUV photodetector (see Section~\ref{sec:detector}). The detector signals are amplified and processed using a digital oscilloscope (Teledyne LeCroy, WavePro 760Zi, 6\,GHz oscilloscope), transferred to a computer, and recorded as a function of the wave number $\tilde\nu_2$.

The fundamental wavenumber of the second ring dye laser is calibrated by recording a laser-induced fluorescence (LIF) spectrum of molecular iodine and the transmission signals through two \'{e}talons using a fraction of the cw laser output. The output of the frequency-fixed first ring dye laser ($\tilde\nu_1=\tfrac13\tilde\nu_{\mathrm{UV}}$) is diffracted using an accousto-optical modulator operated at 675\,MHz. In the case of sum-frequency mixing in Kr, the zero-order beam is transmitted to the amplification chain whereas the frequency of the first-order sideband is locked to the ``t'' hyperfine component of the B $^3\Pi_{0\mathrm u}^+$ ($v'=8$, $J'=129$) $\leftarrow$ X $^1\Sigma_{\mathrm g}^+$ ($v''=4$, $J''=128$) transition of molecular iodine at $15\,682.1648$\,cm$^{-1}$ (Ref. \onlinecite{knoeckel04a}) so that the wave number $2\tilde\nu_{\mathrm{UV}}=6\tilde\nu_1=94\,092.906$\,cm$^{-1}$ ((4p)$^5$ 5p[1/2]$_0$ $\leftarrow$ (4p)$^6$ ${}^1$S$_0$) is located within the bandwidth of the two-photon resonance in Kr.\cite{saloman07a,hollenstein01a} In the case of sum-frequency mixing in Xe, the literature value of the resonance $2\tilde\nu_1= 89\,860.015$\,cm$^{-1}$ ((5p)$^5$ 6p$'$[1/2]$_0$ $\leftarrow$ (5p)$^6$ ${}^1$S$_0$) is used for the calibration.\cite{humphreys70a} The calibration uncertainties of the VUV wave numbers are $0.015$\,cm$^{-1}$ for experiments carried out with Kr as nonlinear gas and about 0.15\,cm$^{-1}$ for those carried out with Xe, which includes the uncertainties caused by the ac Stark shift.

The probe gases (Kr: purity grade 4.0; Ar: 4.8; N$_2$: 5.0; all from Pangas) were used without further purification.

\subsection{Modulation/Demodulation Units}\label{sec:mod-demod}
To exploit the advantages of modulation techniques in experiments with nanosecond laser pulses, the modulation frequency has to be chosen carefully. It has to be higher than the inverse of the duration of a single laser pulse so that there are at least a few oscillation cycles within the typical laser-pulse duration of 2\,ns to 2.5\,ns. The argument can also be formulated in the frequency domain: 
The modulation frequency should exceed the bandwidth of the unmodulated pulse. 
The modulation frequency also has to be low enough so that the bandwidth of the VUV detector (see Section \ref{sec:detector}) can resolve the modulation of the laser pulse resulting from absorption. We found a modulation of about 1.4\,GHz to be an optimal compromise and, therefore, used a resonantly driven electro-optical modulator (EOM).

The fixed output of the modulation source (Anapico APSIN6010) is split into two parts using a coupler (Mini Circuit ZADC-6-2G). The main output of the coupler feeds the reference arm of the demodulator used to demodulate the VUV signal, whereas the coupled output is used to drive the EOM (Qubig T1500M3-400/800). To adjust the intensity of the signal used to modulate the cw output of the second ring dye laser to a desired modulation index $\beta$, a step attenuator (Narda 4748-69) is used in combination with a power amplifier (Becker AMP 20280035). The resonance frequency of the EOM is optimized by minimizing the reflected modulation intensity monitored using a circulator (RYT 300010) and a home-built diode detector and determined to be 1.3875\,GHz. Small variations of the EOM resonance frequency were observed and the necessary adjustments carried out daily. 

To demodulate the VUV signal, we investigated both the standard demodulation technique at $\nu_{\mathrm{mod}}$ (referred to as 1f demodulation below) and the demodulation technique at $2\nu_{\mathrm{mod}}$ (2f) used by Janik {\it et al.} \cite{janik85a}. For 2f detection the modulation signal is amplified (TronTech P23GA) and frequency doubled (Watkins Johnson FD25HC) to drive the demodulation mixer. A small attenuator improves the impedance matching.

In the case of 1f (2f) demodulation, the VUV signal is first amplified (INA 34063) and then split using a diplexer (1f: ALRCOM FC 8312A; 2f: Microwave Circuit D9002G61) into a low- and a high-frequency component. The low-frequency component below 1.05\,GHz (2.0\,GHz) is used to monitor the envelope of the VUV signal and its amplitude represents the transmission signal. The high-frequency component of the VUV signal from the diplexer in the range 1.2--5.0\,GHz (2.4--2.9\,GHz) is demodulated at 1.3875\,GHz (or 2.775\,GHz) using a double balanced mixer (Watkins Johnson WJ-M1G). The demodulated signal is then amplified (Q-Bit QBH-9-131) and recorded with the digital oscilloscope. The phase shift between the VUV signal and the reference signal is adjusted using a ``trombone'' phase shifter (Spinner, 152254) in the reference signal path. On a strong absorption line, the position of the phase shifter for absorption (in-phase, $\phi=0$) and dispersion ($\phi=\tfrac\pi2$) are determined by comparison with calculated line shapes. 
\begin{figure}\centering
	\includegraphics[scale=1]{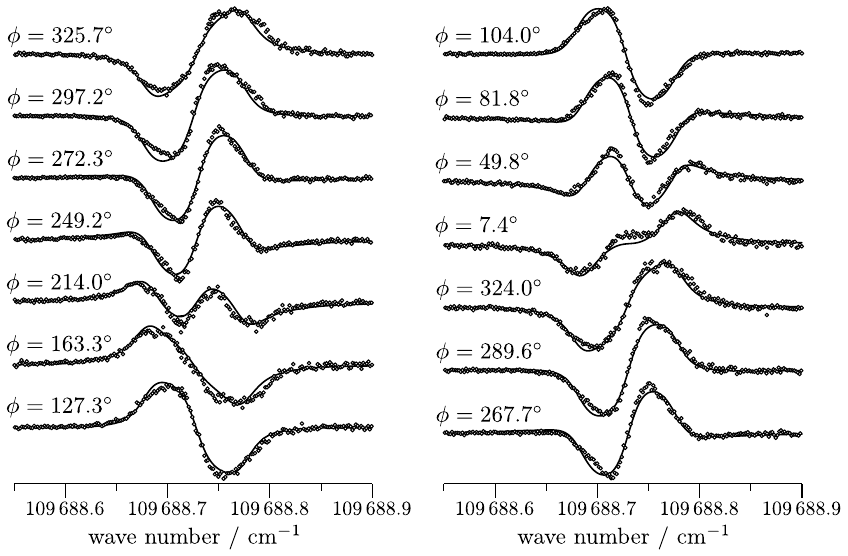}%
	\caption{1f demodulated VUV absorption spectra of the (4p)$^5$ ($^2$P$_{3/2}$) 7d[3/2]$_1$ $\leftarrow$ (4p)$^6$ $^1$S$_0$ transition of krypton measured at a modulation index $\beta=0.5$ for different lengths of the phase shifter. The length of the phase shifter was increased by approximately 2\,cm for each successive trace. 
	The spectra were recorded using only the skimmed supersonic beam in the first experimental chamber. 
	\label{fig:phase}}%
\end{figure}%
As illustration, \figurename~\ref{fig:phase} displays FM spectra of the (4p)$^5$ ($^2$P$_{3/2}$) 7d[3/2]$_1$ $\leftarrow$ (4p)$^6$ $^1$S$_0$ transition of Kr recorded for different lengths of the phase shifter and compares them with calculated spectra for the phases indicated in the respective panels. The lengths of the phase shifter (medium: air) needed to achieve phase shifts of $2\pi$ for the modulation frequency 1.3875\,GHz (2.775\,GHz) is 21.6\,cm (10.8\,cm), which corresponds to the wavelength in ambient air. 

\subsection{VUV Detector}\label{sec:detector}
\begin{figure}\centering
	\includegraphics[scale=1.02]{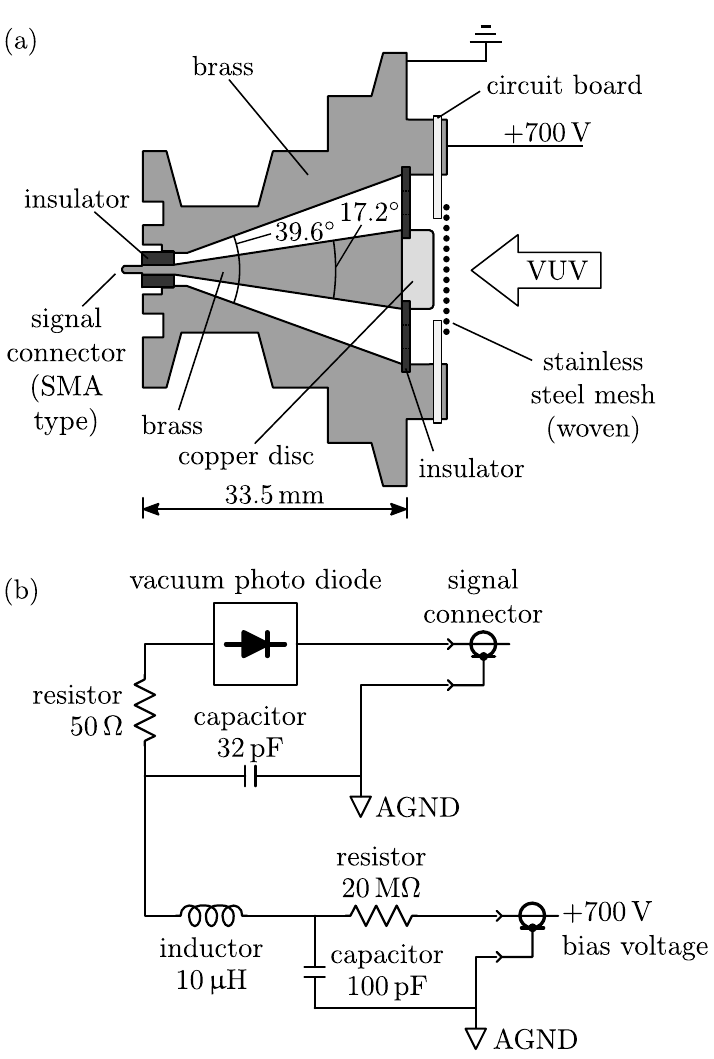}%
	\caption{(a) Section through the cylindrically symmetric home-built VUV photodetector. (b) Corresponding electronic-circuit diagram (see text for details).  
		\label{fig:detector}}%
\end{figure}%
For the experiments presented in this article, we developed a fast vacuum photodiode following principles found in the literature.\cite{tirsell82a,beck76a,green76a} \figurename~\ref{fig:detector} (a) depicts a section through the 10-mm-diameter cylindrically symmetric detector. This home-built detector consists of (i) a 10-mm-diameter polished copper disk serving as photocathode,\cite{cairns66a} (ii) a positively biased anode consisting of a woven stainless-steel mesh with a transmittance of 84\% located at a distance of 1\,mm from the cathode, (iii) a coaxially tapered transmission line with a brass cone as central conductor and an aluminum ISO-KF reduction flange as an outer conductor at ground potential. At the vacuum interface, the diameter of the cathode is reduced to the dimension of the coaxial signal connector with a characteristic impedance of 50\,$\Omega$.\cite{meinke68a} \figurename~\ref{fig:detector} (b) shows a schematic circuit diagram containing most critical components. The 20\,M$\Omega$ resistor, the 100\,pF capacitor and the 10\,$\upmu$H inductor are discrete components and filter the noise from the bias power supply. The capacitance of 32\,pF results from the dielectric properties of the copper-cladded circuit board (Rodgers 3020) holding the anode mesh and is used as charge storage for the signal-induced current. The 50\,$\Omega$ resistor serves as damping element and is composed by 10 resistors of 500\,$\Omega$ soldered in parallel onto the circuit board. The temporal response is governed by circuit properties. The 0--100\% rise time corresponds to the transit time of the photoelectrons (0.1\,ns under typical operation conditions), which is proportional to the distance between cathode and anode and inversely proportional to the square root of the bias voltage. The signal decays exponentially with a time constant ($\tau=120$\,ps) given by the load resistance and the capacitance between cathode and anode.\cite{jordan85a} 

\begin{figure}\centering
	\includegraphics[scale=0.45]{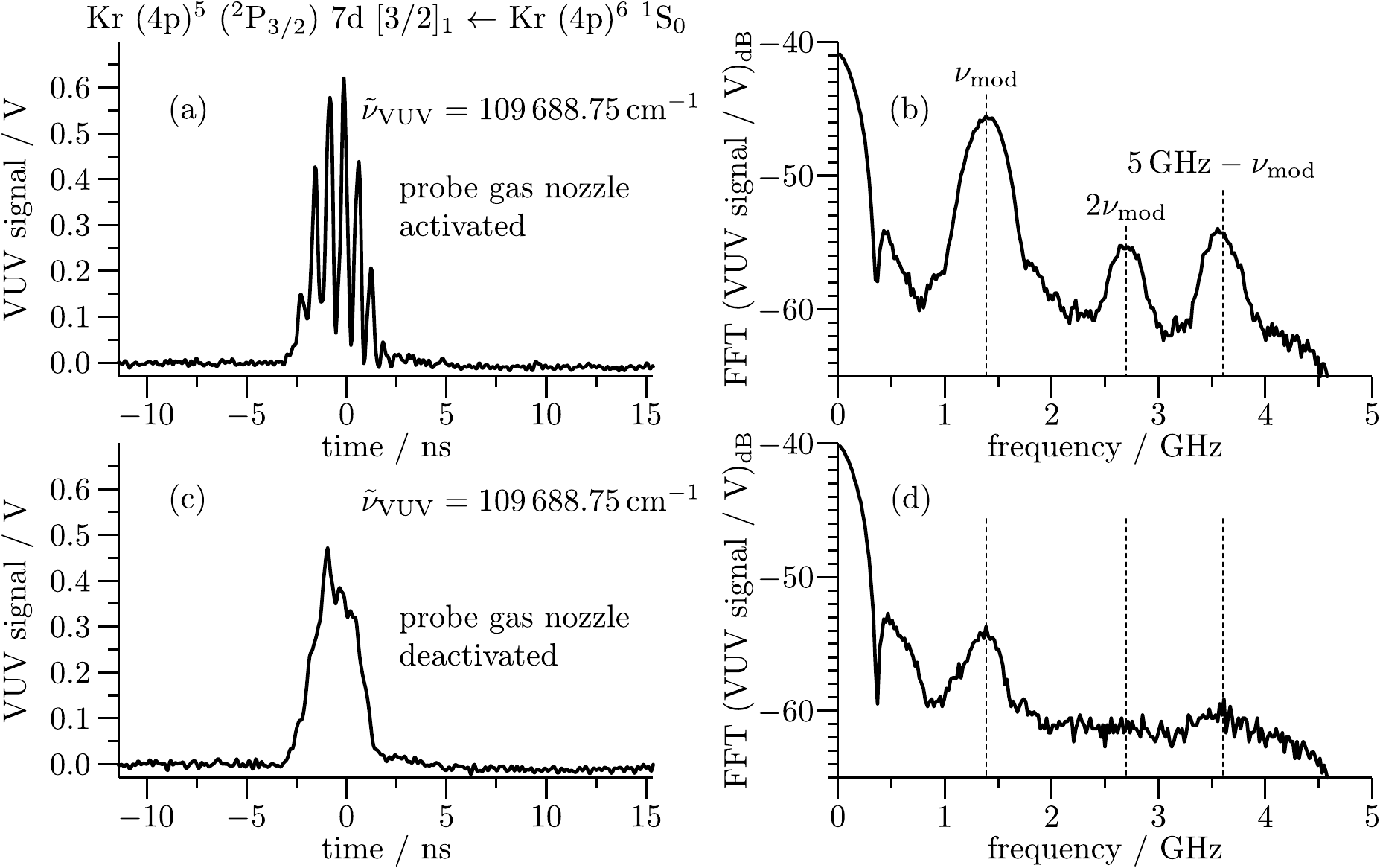}%
	\caption{Time profiles of the frequency-modulated VUV laser pulse recorded under conditions where the lower first-order sideband was resonant with the (4p)$^5$ ($^2$P$_{3/2}$) 7d [3/2]$_1$ $\leftarrow$ (4p)$^6$ $^1$S$_0$ transition in krypton with probe gas nozzle (a) on and (c) off, and corresponding power spectra showing maxima at $\nu_{\mathrm{mod}}$ and $2\nu_{\mathrm{mod}}$ ((c) and (d)). The measurements were carried out at $\nu_{\mathrm{mod}}=1.3875$\,GHz and a modulation index of $\beta\approx0.5$. 
	The widths of the observed bands correspond to the Fourier-transform limit of the envelope of the 2.2-ns-long laser pulse. The band at $5\,\mathrm{GHz}-\nu_{\mathrm{mod}}$ corresponds to an artefact of the digital oscilloscope.\label{fig:mod-fft}}%
\end{figure}%

Lacking an appropriate impulse source, we could not directly measure the time resolution of the detector but instead measured the reflection coefficient up to 6\,GHz, which should ideally be 1.0 for all frequencies. The first deviation occurred at 5.1\,GHz, where it dropped to 0.5. No resonances were detected in the vicinity of the modulation frequency ($\nu_{\mathrm{mod}}$) and its second harmonic ($2\nu_{\mathrm{mod}}$). \figurename~\ref{fig:mod-fft} compares a time trace of the VUV-laser intensity with the laser frequency adjusted such that the lower sideband is resonant with the (4p)$^5$ ($^2$P$_{3/2}$) 7d [3/2]$_1$ $\leftarrow$ (4p)$^6$ $^1$S$_0$ transition in atomic krypton (trace (a)) and a time trace recorded without probe gas (trace (c)). The strong modulation of the VUV intensity resulting from the resonant absorption, with maxima separated by the inverse modulation frequency (0.71\,ns), is clearly visible in \figurename~\ref{fig:mod-fft}(a) and illustrates the high temporal resolution of our home-built VUV detector. The corresponding power spectra obtained by Fourier transformation are depicted in \figurename~\ref{fig:mod-fft}(b) and (d), respectively. The weak residual modulation signal in trace (d) originates from \'etalon effects in the optical setup and absorption by the background gas in the vacuum chamber.

\subsection{Signal detection}\label{sec:signaldetection}
To obtain the VUV FM spectra we used the phase-sensitive detection method rather than the Fourier power spectrum of the time domain VUV-detector signal, which can be easily calculated on the oscilloscope, but is phase-insensitive. The power spectra nevertheless proved to be a useful diagnostic tool to optimize the signals. The time traces of the demodulated signals recorded at the digital oscilloscope and typically averaged over 50 laser shots were transferred to a computer and recorded as a function of the VUV wave number. 

\begin{figure}\centering
	\includegraphics[scale=1.02]{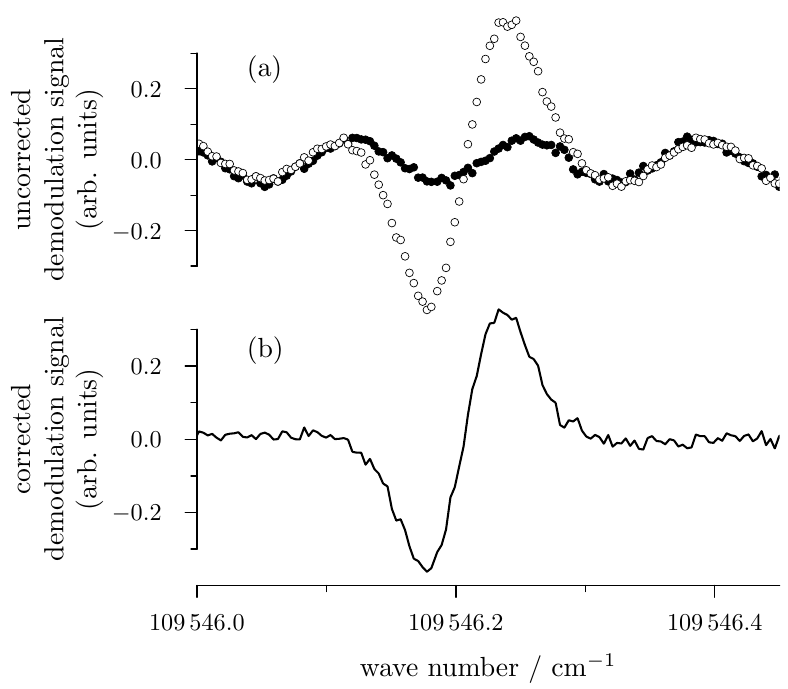}%
	\caption{VUV FM spectrum of the R(0) transition of the N$_2$ b$'$ $^1\Sigma_{\mathrm u}^+$ ($v'=8$) $\leftarrow$ N$_2$ X $^1\Sigma_{\mathrm g}^+$ ($v''=0$) band of molecular nitrogen recorded using all probe gas nozzles. (a) Experimental data points obtained for an optimal temporal overlap of the laser and gas pulse (open circles) and by firing the laser before the gas pulse (full circles). (b) Background-corrected signal, in which contributions of the background gas in the vacuum chamber and \'{e}talon effects are eliminated.\label{fig:cold-warm}}%
\end{figure}%

To obtain artefact-free spectra, it was necessary to remove the background contribution to the signal, as illustrated be \figurename~\ref{fig:cold-warm}, which shows FM VUV spectra of the b$'$ $^1\Sigma_{\mathrm u}^+$ ($v'=8$, $J'=1$) $\leftarrow$ X $^1\Sigma_{\mathrm g}^+$ ($v''=0$, $J''=0$) transition of N$_2$ at 109\,546.2\,cm$^{-1}$ recorded using all probe-gas beams. The top panel corresponds to the signal obtained when the laser is triggered during the gas pulse (open circles) and before the gas pulse (background, full circles) to maintain the same gas load and thus the same background signal in both cases. The latter spectrum reveals a regular oscillation which originates from an \'{e}talon effect affecting the laser beam of frequency $\nu_2$ and also a broad contribution (not seen on the scale of the figure) from the room-temperature background N$_2$ gas in the chamber. The \'{e}talon effect is also noticeable in the actual spectrum. The optimal way to remove the \'{e}talon effect and the background-gas contribution is by measuring the main and background signal in alternating sequences at each frequency step and taking their difference.

\subsection{Modelling the lineshapes}\label{sec:analysis}

To analyse the VUV FM spectra, we followed the approach used by Janik {\it et al.} \cite{janik85a} in their measurements of the sodium D line using 2f FM spectroscopy. The signal contribution at the $k$-th harmonic of the modulation frequency $\omega_{\mathrm{mod}}$ (1.3875\,GHz) is
\begin{alignat}{99}
	S_k = \sum_{n}&\;J_n(\beta)&&J_{n-k}(\beta)\,\times\nonumber \\
	&\exp\left[\vphantom{\frac{\varGamma^2}{2}}\right.&&-\frac{\varGamma^2}{2}\left(\frac{1}{4\varDelta_n^2+\varGamma^2}+\frac{1}{4\varDelta_{n-k}^2+\varGamma^2}\right)
	\nonumber \\ 
	&&&\left.+\mathrm i\varGamma\left(\frac{\varDelta_n}{4\varDelta_n^2+\varGamma^2}-\frac{\varDelta_{n-k}}{4\varDelta_{n-k}^2+\varGamma^2}\right)\right], \label{eq:lineprofile}
\end{alignat}
where $\beta$ is the modulation index (0.5 in the 1f case ($k=1$), 1.25 in the 2f case ($k=2$)), $J_j(\beta)$ are the Bessel functions of order $j$,  $\varDelta_j=\omega+j\omega_{\mathrm{mod}}-\omega_0$, and $\omega$ and $\omega_0$ are the laser and transition frequencies, respectively. In the absence of any additional broadening effect, $S_k$ therefore represents the line profile of an atomic or molecular transition with band center $\omega_0$ and natural linewidth $\varGamma$.

To include the effects of the laser linewidth and Doppler broadening, the calculated line profiles are convoluted with Gaussian functions of appropriate widths. The widths of the Gaussian functions needed to accurately describe the measured line shape were found to depend on the number of pulsed valves used, which points at the fact that the propagation directions of the different supersonic beams are not exactly parallel.

\section{Results}\label{sec:results}
\begin{figure}\centering
	\includegraphics[scale=0.92]{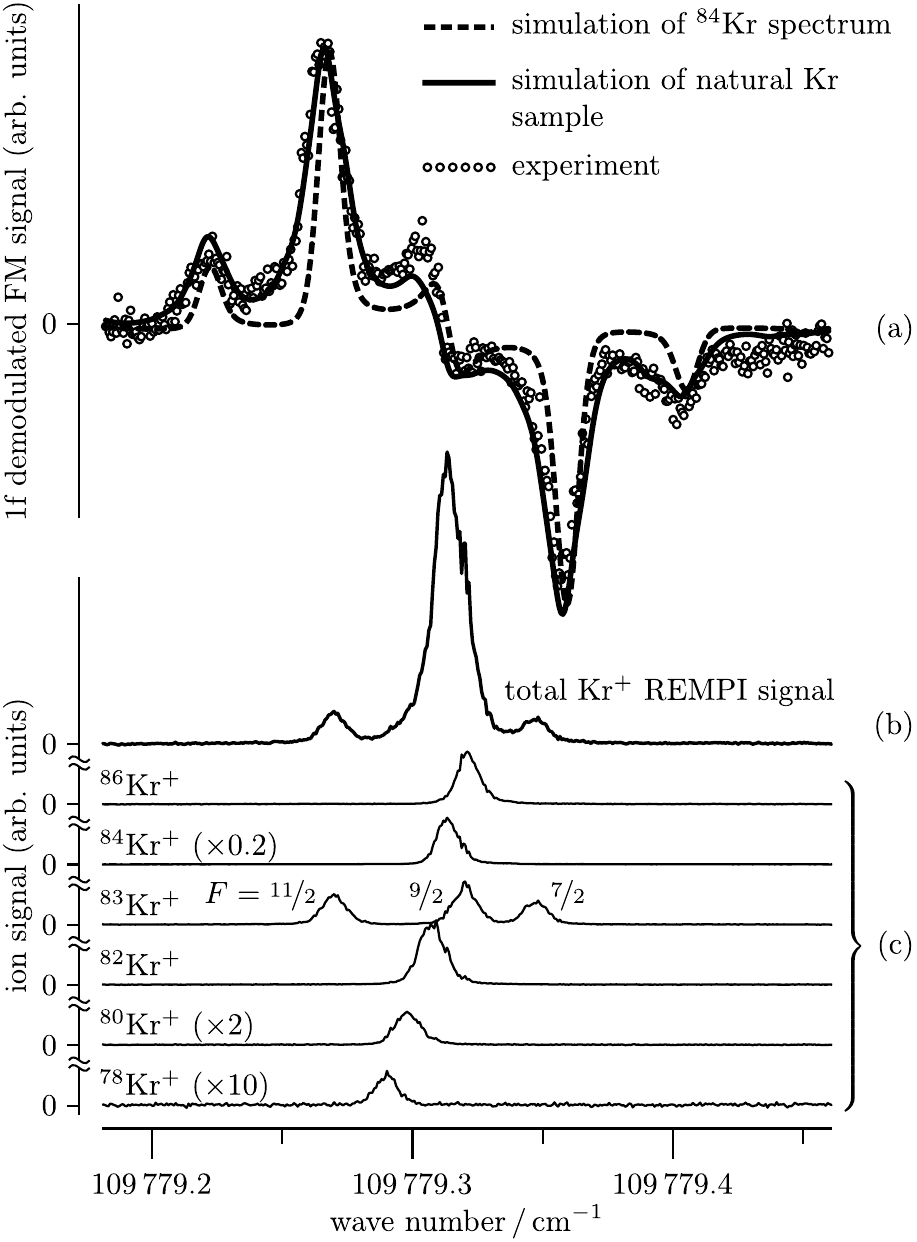}%
	\caption{(a) VUV FM spectrum of the (4p)$^5$ ($^2$P$_{3/2}$) 9s [3/2]$_1$ $\leftarrow$ (4p)$^6$ $^1$S$_0$ transition of krypton recorded with modulation index of 1.25 and 1f demodulation at $\phi=190^{\circ}$ (open circles). The thick full and dashed lines are the corresponding simulated FM spectra (1f, $\beta=1.25$, $\phi=190^{\circ}$) for a pure ${}^{84}$Kr sample and a natural Kr sample, respectively. (b) ($1+1'$) REMPI spectrum of a natural Kr gas sample obtained by summing the spectra of the different isotopes in (c). The ($1+1'$) REMPI spectra were recorded without modulating the VUV frequency. The vertical scale is linear.
	\label{fig:kr-7d}}%
\end{figure}%
\begin{table}
\caption{\label{tab:kr:1} Wave numbers of the (4p)$^5$ ($^2$P$_{3/2}$) 9s [3/2]$_1$ $\leftarrow$ (4p)$^6$ $^1$S$_0$ transition of the natural isotopes of Kr. 
}
\begin{ruledtabular}
	\begin{tabular}{lc}
		Isotope & $\frac{\tilde\nu^{\mathrm{REMPI}}}{\mathrm{cm^{-1}}}$ \\ \hline
		${}^{78}$Kr           & 109\,779.290(16)  \\
		${}^{80}$Kr           & 109\,779.298(16)  \\
		${}^{82}$Kr           & 109\,779.307(16)  \\
		${}^{83}$Kr ($F=11/2$)& 109\,779.270(16)  \\
		${}^{83}$Kr ($F=9/2$) & 109\,779.320(16)  \\
		${}^{83}$Kr ($F=7/2$) & 109\,779.347(16)  \\
		${}^{84}$Kr           & 109\,779.314(16)  \\
		${}^{86}$Kr           & 109\,779.321(16)  \\ 
	\end{tabular}
\end{ruledtabular}
\end{table}
\subsection{Krypton}\label{sec:results:kr}
\begin{figure*}\centering
	\includegraphics[scale=1.1]{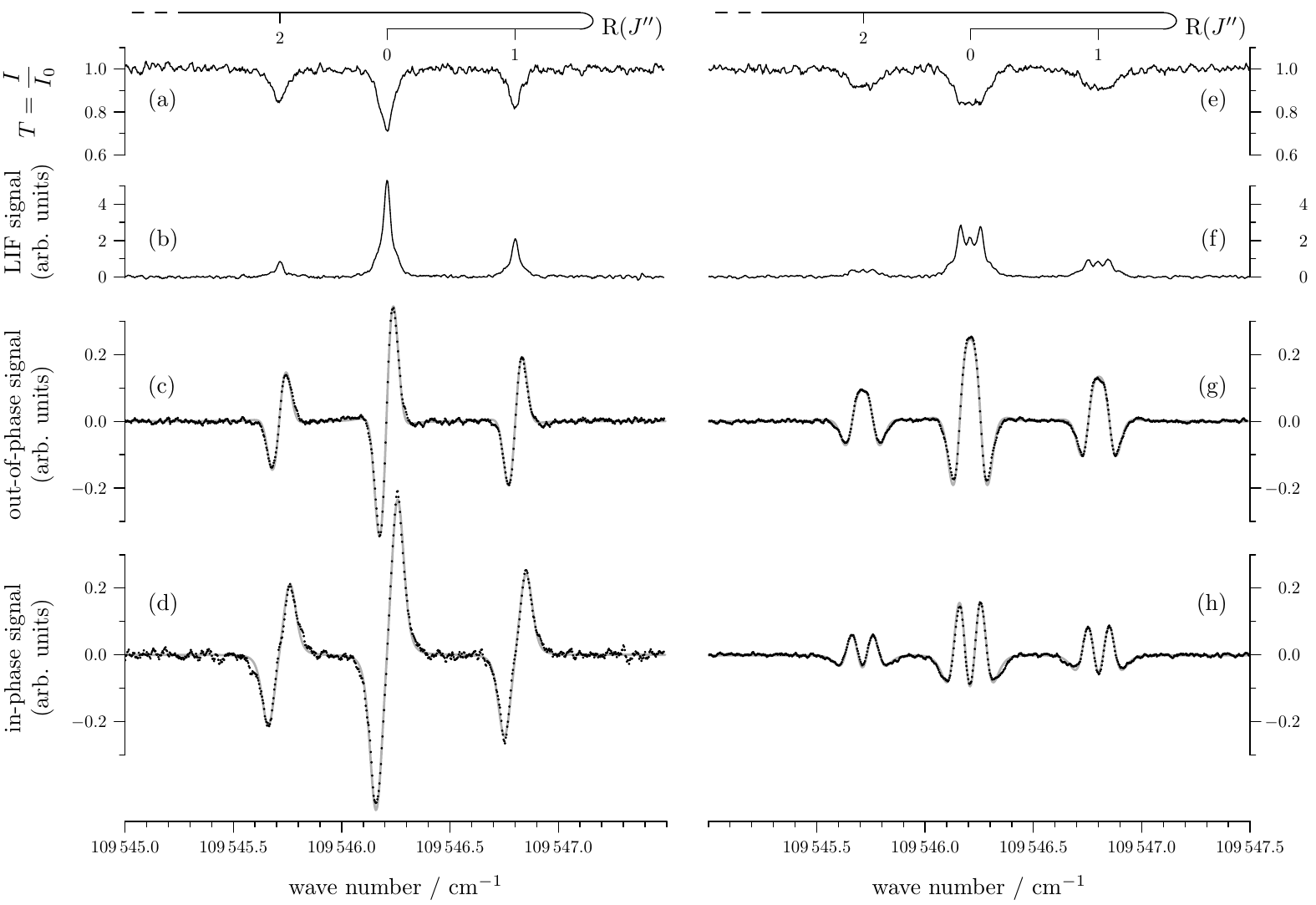}%
	\caption{VUV photoabsorption spectrum of the R-branch (band head) of the N$_2$ b$'$ $^1\Sigma_{\mathrm u}^+$ ($v'=8$) $\leftarrow$ N$_2$ X $^1\Sigma_{\mathrm g}^+$ ($v''=0$) transition in molecular nitrogen recorded using all probe gas nozzles. The top traces ((a) and (e)) are the transmittance spectra, the traces (b) and (f) the LIF spectra, the traces (c) and (g) the demodulated out-of-phase FM spectra ($\phi\approx\frac\pi2$) and the traces (d) and (h) the demodulated in-phase FM spectra ($\phi\approx 0$). The modulation indices $\beta$ used to record (a)--(d) and (e)--(h) were $\beta=0.5$ and 1.25 corresponding to 1f and 2f demodulation, respectively. The dots in traces (c), (d), (g) and (h) are experimental data point, whereas the solid line is the fitted spectrum using Equation~\eqref{eq:lineprofile}.\label{fig:n2:1f2f:comp}}%
\end{figure*}%
Panel (a) in \figurename~\ref{fig:kr-7d} shows the FM spectrum (open circles) of the (4p)$^5$ ($^2$P$_{3/2}$) 9s [3/2]$_1$ $\leftarrow$ (4p)$^6$ $^1$S$_0$ transition of Kr recorded with a modulation index of $\beta=1.25$ and a 1f demodulation phase of 190$^{\circ}$. The spectrum was obtained using only the supersonic beam in the photoionization chamber. For comparison, ($1+1'$) REMPI spectra of the natural isotopes of Kr are depicted in panel (c) of the figure. The ion signals from the different isotopes were separated using a linear time-of-flight mass spectrometer as already demonstrated in Ref. \onlinecite{hollenstein03a}.
The frequency-doubled output (532\,nm) of the Nd:YAG laser was used in the ionization step ($1'$). These spectra were measured without FM and clearly reveal the expected blue-shift of the transition frequency with increasing mass of the $I=0$ isotopes ($^{78}$Kr, $^{80}$Kr, $^{82}$Kr, $^{84}$Kr, $^{86}$Kr) \cite{brandi02a} and the hyperfine structure of the $^{83}$Kr isotope ($I=\frac92$). Summing the REMPI spectra of all isotopes leads to the spectrum depicted in \figurename~\ref{fig:kr-7d}(b). The central line in this spectrum, with a full width at half maximum of 0.015\,cm$^{-1}$ and a maximum at the position of the dominant $^{84}$Kr isotope, is slightly asymmetric because of the contributions of the other isotopes. The two weak satellite lines with full width at half maximum of 0.011\,cm$^{-1}$, primarily given by the laser linewidth and the Doppler broadening, originate exclusively from the $^{83}$Kr isotope. The thick full line in the top panel in \figurename~\ref{fig:kr-7d} represents a simulation of the FM spectrum obtained numerically from the REMPI spectrum in panel (b) for $\beta=1.25$ and $\phi=190^{\circ}$. The dashed line in \figurename~\ref{fig:kr-7d}(a) corresponds to a numerical simulation of the FM of the spectrum of the $^{84}$Kr isotope only. The comparison of the experimental with the two simulated spectra clearly indicates that the FM spectrum of the isotopic mixture contains information on the isotope shifts and hyperfine structure although this information is not as easy to recognize as in the REMPI spectra. The analysis of all spectra presented in \figurename~\ref{fig:kr-7d} leads to the spectral positions listed in \tablename~\ref{tab:kr:1}, which are fully consistent with the term values recommended in Ref. \onlinecite{saloman07a}, and to the conclusion that the natural (Lorentzian) line profile of the transition and the Gaussian profile resulting from the laser profile and the Doppler effect have full widths at half maximum of 0.0028(10)\,cm$^{-1}$ and 0.0087(5)\,cm$^{-1}$, respectively.

\subsection{N$_{\mathbf{2}}$}\label{sec:results:N2}
The ability to simultaneously record LIF and FM spectra in the VUV is demonstrated in \figurename~\ref{fig:n2:1f2f:comp} with the example of the R(0), R(1) and R(2) lines of the b$'$ $^1\Sigma_{\mathrm u}^+$ ($v'=8$) $\leftarrow$ X $^1\Sigma_{\mathrm g}^+$ ($v''=0$) transition of N$_2$. The spectra displayed in the left and right panels of the figure were measured with a modulation index $\beta$ of 0.5 and 1.25, respectively. In each panel, the traces correspond, from top to bottom, to the transmission spectra through the eleven supersonic beams of pure N$_2$ gas in the photoionization and absorption chambers ((a) and (e)), the LIF spectra collected on the MCP detector located in the photoionization chamber ((b) and (f)) and the FM spectra recorded using demodulation phases of $\frac\pi2$ ((c) and (g)) and 0 ((d) and (h)), respectively.

At the nozzle stagnation pressure of 2\,bar used in the experiments, the eleven skimmed supersonic beams absorb about 25\% of the VUV radiation at the position of the strongest (R(0)) line. Under these conditions, the signal-to-noise ratio of the transmission spectrum ($\approx$10) is limited by the shot-to-shot-fluctuations of the VUV-laser intensity. The lineshapes are determined by the sideband structure generated at the two different modulation indices. The sideband structure is not resolved in the transmission spectrum because of the Doppler broadening of about 1.5\,GHz (FWHM) resulting from the eleven beams. Consequently, the lines recorded at $\beta=1.25$ have broad, flat-top lineshapes and are about three times broader than the lines recorded at $\beta=0.5$, in which the weaker sidebands manifest themselves as shoulders on both sides of the lines. The sideband structures are more clearly observed in the LIF spectra because the single beam in the photoionization chamber only leads to a Doppler broadening of 1\,GHz (FWHM). The signal-to-noise ratio of the LIF spectra is much higher ($\approx$50), despite the low collection efficiency ($\approx$0.7\%) of the fluorescence imposed by the small solid-angle of the detection, because of the background-free nature of LIF.

The FM spectra are also characterized by a high signal-to-noise ratio (between 40 and 60) which represents an improvement by a factor of more than 5 over the transmission spectrum. The lineshapes correspond exactly to the lineshapes calculated using Eq.~\eqref{eq:lineprofile} for a Lorentzian linewidth (FWHM) of 0.05(1)\,cm$^{-1}$, after convolution with a Gaussian line profile with FWHM of 0.14\,cm$^{-1}$. The highest signal-to-noise ratio ($\approx$60) is obtained by 2f modulation in combination with a demodulation phase of $\frac\pi2$. The positions of the three transitions derived from the spectra presented in \figurename~\ref{fig:n2:1f2f:comp} are listed in the upper part of \tablename~\ref{tab:n2:1}.
\begin{table}
	\caption{\label{tab:n2:1} Wave numbers of the b$'$ $^1\Sigma_{\mathrm u}^+$ ($v'=8$) $\leftarrow$ X $^1\Sigma_{\mathrm g}^+$ ($v''=0$) R-branch transitions and of the rovibrational lines of the b $^1\Pi_{\mathrm u}$ ($v'=12$) $\leftarrow$ X $^1\Sigma_{\mathrm g}^+$ ($v''=0$) transition of molecular nitrogen. The wave numbers in the third and fourth columns have uncertainties of 0.016\,cm$^{-1}$ and $0.3-0.8$\,cm$^{-1}$, respectively.
	}
	\begin{ruledtabular}
		\begin{tabular}{llcc}
			& & this work & Ref.~\onlinecite{carroll69a} \\
			\multicolumn{2}{l}{Transition} & $\tilde\nu\,/\,\mathrm{cm^{-1}}$ & $\tilde\nu\,/\,\mathrm{cm^{-1}}$\\ \hline
			b$'$(8)$\leftarrow$X(0)&R(0) & 109\,546.206 \\
			&R(1) & 109\,546.800 \\
			&R(2) & 109\,545.710 \\ \hline 
			b(12)$\leftarrow$X(0)& 
			R(0) & 109\,833.454 & 109\,833.5\footnote{\footnotemark[2]~\footnotemark[3]~\footnotemark[4]~Unresolved transitions.} \\
			&R(1) & 109\,834.071 & 109\,833.5\footnotemark[1] \\
			&R(2) & 109\,833.003 & 109\,833.5\footnotemark[1] \\
			&R(3) & 109\,830.257 & 109\,829.9\footnotemark[2] \\
			&R(4) & 109\,825.826 & 109\,825.9\footnotemark[3] \\
			&Q(1) & 109\,829.475 & 109\,829.9\footnotemark[2] \\
			&Q(2) & 109\,826.112 & 109\,825.9\footnotemark[3] \\
			&Q(3) & 109\,821.065 & 109\,821.1\footnotemark[4] \\
			&P(2) & 109\,821.520 & 109\,821.1\footnotemark[4] \\
		\end{tabular}
	\end{ruledtabular}
\end{table}
\begin{figure*}\centering
	\includegraphics[scale=1.05]{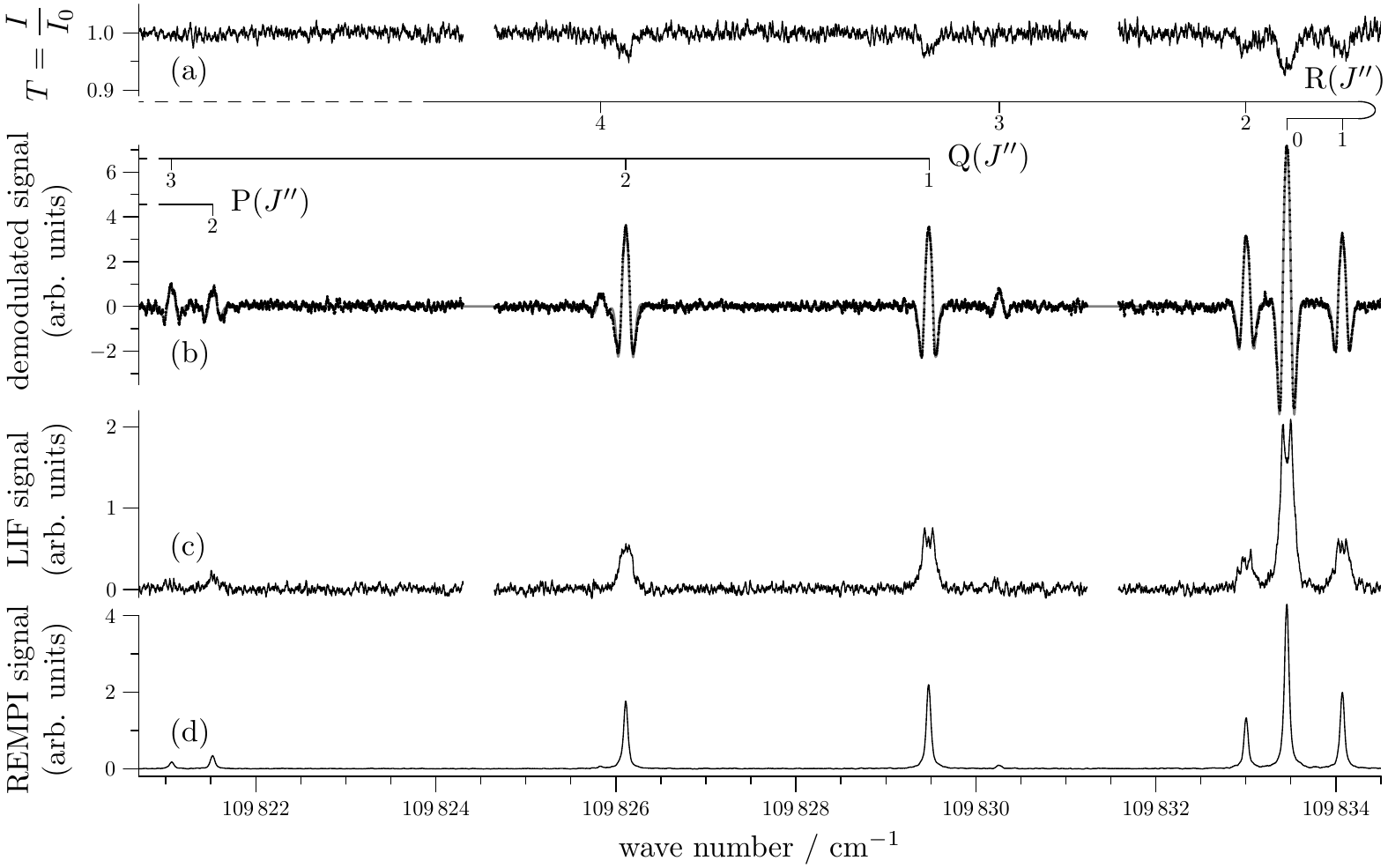}%
	\caption{VUV spectra of the N$_2$ b $^1\Pi_{\mathrm u}$ ($v'=12$) $\leftarrow$ N$_2$ X $^1\Sigma_{\mathrm g}^+$ ($v''=0$) transition of molecular nitrogen. The three top traces traces show the transmittance (a), the 2f-demodulated FM (b), and the LIF (c) spectra recorded using all probe-gas beams and a modulation index of $\beta\approx1.25$. The bottom trace is a ($1+1'$) REMPI spectrum recorded without modulating the VUV-laser frequency (with one valve only). \label{fig:n2:2f}}%
\end{figure*}%
Transmission, FM, and LIF spectra of the b $^1\Pi_{\mathrm u}$ ($v'=12$) $\leftarrow$ X $^1\Sigma_{\mathrm g}^+$ ($v''=0$) transition of molecular nitrogen are compared in \figurename~\ref{fig:n2:2f}(a)-(c). They were obtained using a modulation index $\beta=1.25$, a demodulation phase of $\frac\pi2$ and all eleven supersonic beams. The lines have the characteristic shape discussed above. The signal-to-noise ratio of the transmission spectrum is only about 3 for the strongest line (R(0)) and insufficient to observe the weak P(2), Q(3), R(3) and R(4) lines. These lines, however, are clearly observed in the 2f FM spectrum, which even has a better signal-to-noise ratio ($\approx$20 for R(0)) than the LIF spectrum. \figurename~\ref{fig:n2:2f}(d) is a ($1+1'$) REMPI spectrum recorded without modulating the VUV laser frequency and using only the beam in the photoionization chamber. This spectrum has by far the best signal-to-noise ratio ($>$300 for R(0)) and its lines are characterized by Lorentzian lineshapes with FWHM of 0.05(1)\,cm$^{-1}$, corresponding to a lifetime of about $106(21)$\,ps, which we attribute to predissociation. This lifetime is shorter than the laser pulses, which explains the reduced signal-to-noise ratio of the LIF spectrum compared to that shown in \figurename~\ref{fig:n2:1f2f:comp}.

All transition wave numbers of the b(12) $\leftarrow$ X(0) lines are listed in the lower part of \tablename~\ref{tab:n2:1} where they are compared with the earlier results of Refs.~\onlinecite{carroll69a}.

\subsection{Argon}
\begin{figure}\centering
	\includegraphics[scale=1]{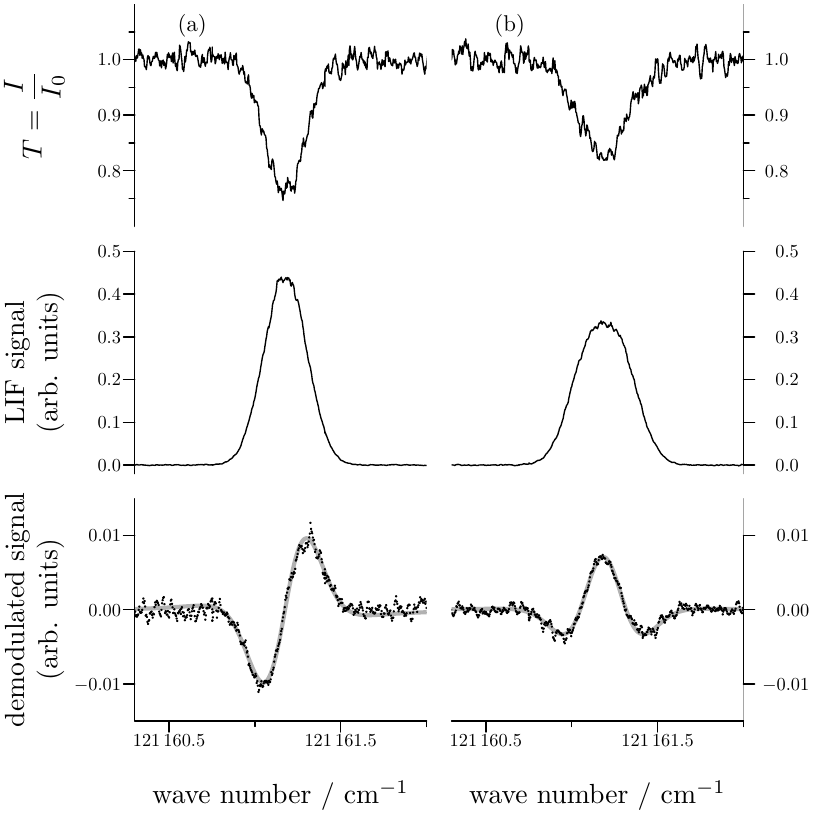}%
	\caption{VUV photoabsorption spectrum of the (3p)$^5$ 6s$'$ [1/2]$_1$ $\leftarrow$ (3p)$^6$ $^1$S$_0$ transition of argon. The top, middle and bottom traces show the transmittance, the LIF and the demodulated FM spectra, respectively. On the left-hand-side (right-hand-side) of the figure the VUV signal was demodulated using 1f-demodulation (2f-demodulation) using a modulation index of $\beta\approx0.5$ ($\beta\approx1.25$) and a demodulation phase $\phi=\frac\pi2$. The spectra were recorded using the first probe-gas nozzle only. \label{fig:ar:1f2f}}%
\end{figure}%
Transmission, LIF and FM spectra of the (3p)$^5$ 6s$'$ [1/2]$_1$ $\leftarrow$ (3p)$^6$ $^1$S$_0$ transition of argon near 121\,160\,cm$^{-1}$ are presented in \figurename~\ref{fig:ar:1f2f}. They were obtained after frequency doubling the pulsed-amplified modulated output of the second ring dye laser and using only the supersonic beam in the photoionization chamber. The transition is so strong that about 20\% of the VUV radiation is absorbed by a single skimmed supersonic beam and the VUV laser intensity used in the experiments ($10^{10}$\,photons/pulse in an area of $\approx$0.1\,mm$^2$) saturates the strong transitions. The transmission spectra have a signal-to-noise ratio of about 3.

Modulation at $\beta=0.5$ (1f) and 1.25 (2f) followed by demodulation at $\phi=\frac\pi2$ gives a signal-to-noise ratio of about 10. The line center is 121\,161.18(15)\,cm$^{-1}$, which is consistent with the value of 121\,161.3135\,cm$^{-1}$ recommended in Ref. \onlinecite{NIST2016a}.
The observed lineshapes can be exactly reproduced using Eq.~\eqref{eq:lineprofile} and convolution with a Gaussian lineshape function with full width at half maximum of 0.25\,cm$^{-1}$ assuming that the Lorentzian linewidth ($\varGamma$ in Eq.~\eqref{eq:lineprofile}) is 0.3\,cm$^{-1}$, which is much larger than the natural linewidth of the 6s$'$[1/2]$_1$ level because of saturation. The broad lines are also the reason why the signal-to-noise ratio of the FM spectra is only slightly higher than that of the transmission spectrum. Indeed, maximal contrast and sensitivity is reached in FM spectroscopy when the modulation frequency is larger than the linewidth of the transition (see \figurename~\ref{fig:mod-spec}).

\section{Conclusions}\label{sec:conclusions}
In this article, we have presented an extension of FM spectroscopy to the VUV range of the electromagnetic spectrum. This extension enables the recording of absorption spectra in the range between 200\,nm and 60\,nm using table-top pulsed VUV lasers based on resonance-enhanced four-wave mixing. The attractive features of FM spectroscopy in the VUV are (i) its high sensitivity, which results from its background-free nature and the fact that the signal-to-noise ratio is not limited by shot-to-shot fluctuations of the VUV laser intensity; (ii) the sensitivity is sufficiently high that spectra of cold samples in skimmed supersonic beams can be recorded with high signal-to-noise ratio under conditions where no significant attenuation of the VUV radiation can be detected; (iii) the ease with which it can be combined with other detection methods such as LIF, photoionization mass spectrometry and ($1+1'$) REMPI spectroscopy to characterize the photophysical processes resulting from the absorption of short-wavelength radiation, and (iv) its high resolution, which results from the intrinsically narrow bandwidth of Fourier-transform-limited pulsed VUV lasers. Its main drawback is that very broad spectral features or absorption continua cannot be detected as sensitively as sharp spectral features.

To illustrate the principles of FM spectroscopy in the VUV, we have presented absorption, LIF, and ($1+1'$) REMPI spectra of Ar, Kr and N$_2$ near 82.5\,nm and 92\,nm. Optimal sensitivity was reached for modulation frequencies between 1\,GHz and 2\,GHz in combination with a modulation index of 1.25, which maximizes the intensity of the first modulation sidebands, and demodulation at twice the modulation frequency (2f) and a demodulation phase of $\phi=\frac\pi2$. This large modulation index can lead to a broadening of the spectral lines when the linewidths of the transitions exceed the modulation frequency. In this case, it can be more advantageous to use a modulation index of 0.5 in combination with 1f demodulation at $\phi=\frac\pi2$.

\begin{acknowledgments}
This work is supported financially by the Swiss National Science Foundation under the project No. 200020-159848/1.
\end{acknowledgments}


%



\end{document}